\documentclass[prb,twocolumn,showpacs,amsmath,amssymb,preprintnumbers,superscriptaddress]{revtex4}
\usepackage{dcolumn}
\usepackage{bm}
\usepackage{graphicx}
\usepackage{subfigure}
\usepackage{color}

\begin{document}

\title{Effect of Cu$^{2+}$ substitution in Spin-Orbit Coupled Sr$_2$Ir$_{1-x}$Cu$_x$O$_4$: Structure, magnetism and electronic properties}

\author{Imtiaz Noor Bhatti}\affiliation{School of Physical Sciences, Jawaharlal Nehru University, New Delhi - 110067, India.}
\author{R. S. Dhaka}\affiliation{Department of Physics, Indian Institute of Technology Delhi, Hauz Khas, New Delhi - 110016, India.}
\author{A. K. Pramanik}\email{akpramanik@mail.jnu.ac.in}\affiliation{School of Physical Sciences, Jawaharlal Nehru University, New Delhi - 110067, India.}

\begin{abstract}
Sr$_2$IrO$_4$ is an extensively studied spin-orbit coupling induced insulator with antiferromagnetic ground state. The delicate balance between competing energy scales plays crucial role for its low temperature phase, and the route of chemical substitution has often been used to tune these different energy scales. Here, we report an evolution of structural, magnetic and electronic properties in doped Sr$_2$Ir$_{1-x}$Cu$_x$O$_4$ ($x$ $\leq$ 0.2). The substitution of Cu$^{2+}$ (3$d^9$) for Ir$^{4+}$ (5$d^5$) acts for electron doping, though it tunes the related parameters such as, spin-orbit coupling, electron correlation and Ir charge state. Moreover, both Ir$^{4+}$ and Cu$^{2+}$ has single unpaired spin though it occupies different $d$-orbitals. With Cu substitution, system retains its original structural symmetry but the structural parameters show systematic changes. X-ray photoemission spectroscopy measurements show Ir$^{4+}$ equivalently converts to Ir$^{5+}$ and a significant enhancement in the density of states has been observed at the Fermi level due to the contribution from the Cu 3$d$ orbitals, which supports the observed decrease in the resistivity with Cu substitution. While the long-range magnetic ordering is much weakened and the highest doped sample shows almost paramagnetic-like behavior the overall system remains insulator. Analysis of resistivity data shows mode of charge conduction in whole series follows 2-dimensional variable-range-hopping model but the range of validity varies with temperature. Whole series of samples exhibit negative magnetoresistance at low temperature which is considered to be a signature of weak localization effect in spin-orbit coupled system, and its evolution with Cu appears to follow the variation of resistivity with $x$. 
\end{abstract}


\maketitle
\section{Introduction}
In recent times, Ir-based oxides are the most extensively studied materials for its many interesting properties.\cite{rau,krampa,cao, mandru, christianson, wan, yin, dally, kumar, kim1} The layered Sr$_2$IrO$_4$ is of prime interest due to its exotic $J_{eff}$ = 1/2 ground state. The Ir in Sr$_2$IrO$_4$ adopts ionic state of Ir$^{4+}$ which has 5$d^5$ electronic configuration. The crystal field effect (CFE) in IrO$_6$ octaheral environment splits 5$d$ orbital into $t_{2g}$ and $e_g$ levels, then in presence of strong spin-orbit coupling (SOC) effect the low lying t$_{2g}$ level is further split in to fully filled $J_{eff}$ = 3/2 quartet and partially filled $J_{eff}$ = 1/2 doublet state.\cite{kim1, kim2} These $J_{eff}$ = 1/2 pseudospins engage in Heisenberg-type antiferromagnetic (AFM) exchange interaction, however, the rotation/distortion of IrO$_6$ octahedra induces Dzyaloshinskii-Moriya (DM) type antisymmetric interaction which is believed to give rise weak ferromagnetic (FM) behavior in this material.\cite{crawford,ye}

The bilayer nature of Sr$_2$IrO$_4$ is very evident in asymmetric magnetic exchange interaction along in-plane and out-of-plane direction.\cite{fujiyama} Moreover, experimental studies have shown evolution of magnetism is very linked with its structural modifications.\cite{jack,imtiaz} This material is surprisingly an insulator with extended 5$d$ orbitals which provides a low electronic correlation effect ($U$), hence Sr$_2$IrO$_4$ would otherwise be a metal. In addition, following structural symmetry with other 3$d$ and 4$d$ based superconducting (SC) materials i.e. doped La$_2$CuO$_4$,\cite{tarascon,bozin} and Sr$_2$RuO$_4$,\cite{maeno,mack} respectively the recent theoretical studies have predicted possible SC state in doped Sr$_2$IrO$_4$.\cite{yang,gao,casa,yan,sumita} While experimental data have not supported any SC state till date, but the angel resolved photoemission spectroscopy (ARPES) measurements have interestingly shown some exotic electronic states in doped Sr$_2$IrO$_4$ such as, disconnected Fermi arcs and rapid collapse of Mott gap with emergence of nodal excitations where these features are similarly seen in doped cuprates.\cite{ykkim,torre} 

With an aim to understand the exotic magnetic ground state as well as to tune the electronic properties for realization of possible SC behavior, approach of chemical substitution with various elements has recently been adopted both at Sr- and Ir-site. For instance, collapse of long-range magnetic state and evolution of metallic behavior has generally been observed in case of electron doped of (Sr$_{1-x}$La$_x$)IrO$_4$.\cite{chen,gretarsson} Some of the Ir-site doping has yielded similar results. The Rh substitution in Sr$_2$Ir$_{1-x}$Rh$_x$O$_4$ has shown suppression of both ordered magnetic and insulating state where the system evolve to paramagnetic (PM) and metallic state above critical concentration of Rh in system.\cite{clancy,qi} Similarly, Ru substitution has caused suppression of magnetic ordering in Sr$_2$IrO$_4$ with nearly 30 to 50\% of Ru$^{4+}$ doping level.\cite{calder,yuan} Interestingly, recent substitution of 4$f$ element in Sr$_2$Ir$_{1-x}$Tb$_x$O$_4$ shows mere 3\% of Tb$^{4+}$ completely suppresses the AFM transition while retaining its insulating behavior.\cite{wang}

In present work, we have investigated the effect of Cu substitution in Sr$_2$Ir$_{1-x}$Cu$_x$O$_4$ which is rather interesting as Cu$^{2+}$ (3$d^9$) doped for Ir$^{4+}$ (5$d^5$) not only act for electron doping, but simultaneously it will tune both SOC and $U$, however, in opposite manner. Another obvious effect would be the change of Ir charge state from Ir$^{4+}$ to Ir$^{5+}$ (5$d^4$) where later is believed to be nonmagnetic ($J_{eff}$ = 0) and will act for site dilution. Nonetheless, both Ir$^{4+}$ and Cu$^{2+}$ has single unpaired electrons, though they occupy different and orthogonal $t_{2g}$ and $e_g$ orbitals, respectively. This would really be interesting to understand the evolution of magnetic and electronic behavior in Sr$_2$Ir$_{1-x}$Cu$_x$O$_4$ and to test the compatibility of these orbitals in promoting the magnetism and electronic conduction in this system. Our present work is also motivated to realize the recent calculation which shows an evolution of PM but spin-orbital ordered Mott phase, particularly doping with elements having extremely weak SOC character which act as a giant perturbation to the spin-orbital structure of Sr$_2$IrO$_4$.\cite{martins}

Our results show that doped Cu$^{2+}$ substitutes Ir$^{4+}$ and alters the charge state of Ir producing roughly double amount of Ir$^{5+}$. The overall system retains the original structural symmetry, though the lattice parameters and the distortion of IrO$_6$ octahedra modifies with doping of Cu. We observe that with progressive Cu substitution, long-range magnetic state in Sr$_2$IrO$_4$ is destabilized and the system converts to nearly PM state with $\sim$ 20\% of Cu concentration. The resistivity decreases continuously but we have not evidenced metallic state within this studied concentration of Cu. Our analysis shows temperature dependent charge conduction is well explained with Mott's two dimensional (D) variable-range-hopping (VRH) mechanism.

\begin{figure}
	\centering
		\includegraphics[width=8cm, height=12cm]{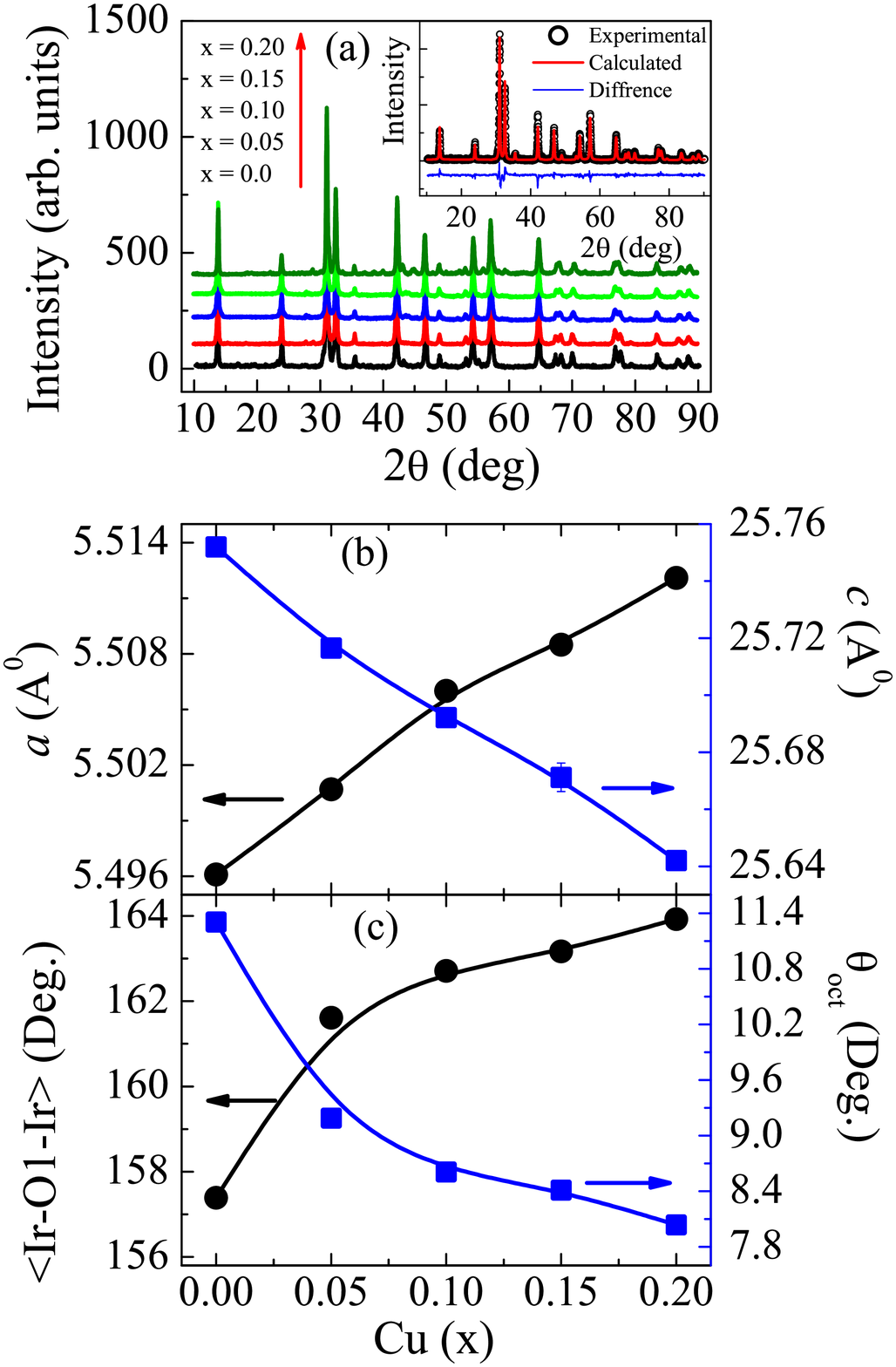}
	\caption{(color online) (a) shows the XRD pattern for Sr$_2$Ir$_{1-x}$Cu$_{x}$O$_4$ series. Inset shows the Rietveld refinement for $x$ = 0.0 sample. (b) shows lattice parameters $a$ (left axis) and $c$ (right axis) and (c) shows bond angle $<$Ir-O1-Ir$>$ (left axis) and octahedral rotation $\theta_{Oct}$ (right axis) as a function of Cu concentration $x$. These parameters are determined from Rietveld analysis of powder x-ray diffraction data.}
	\label{fig:Fig1}
\end{figure}

\section{Experimental Detail}
Series of polycrystalline materials Sr$_2$Ir$_{1-x}$Cu$_x$O$_4$ ($x$ = 0.0, 0.05, 0.10, 0.15 and 0.20) are prepared by conventional solid state method. High purity ingredient components SrCO$_3$, IrO$_2$ and CuO are mixed in stoichiometric ratio and ground well. The mixed powders are then given several heat treatment in powder and pellet form with intermediate grindings. Finally, samples are prepared after giving heat treatment at 1100$^o$C. The phase purity of all the materials has been checked by x-ray diffraction (XRD) where the XRD data have been analyzed with Rietveld refinement program. The XRD data have been collected using a Rigaku made diffractometer. Data are collected in 2$\theta$ range 10 - 90$^o$ at an interval of 0.02$^o$. The details of sample preparation and characterization are reported elsewhere.\cite{imtiaz,imtiaz1} The quantitative analysis of elemental composition of present Sr$_2$Ir$_{1-x}$Cu$_x$O$_4$ series has been done with energy dispersive analysis of x-ray (EDX). The obtained molar concentration of ions in Sr$_2$Ir$_{1-x}$Cu$_x$O$_4$ closely matches with the nominal concentration of ions within error limit of $\pm$ 3-4\%. Further, x-ray photoemission spectroscopy (XPS) study has been done to probe the oxidation state of iridium and copper. The XPS measurements are performed with base pressure in the range of 10$^{-10}$ mbar using a commercial electron energy analyzer (PHOIBOS 150 from Specs GmbH, Germany) and a non-monochromatic Al$K\alpha$ x-ray source ($h\nu$ = 1486.6 eV). The DC magnetization is measured in Physical Properties Measurement System (Quantum Design) whereas electrical transport measurement is done on home made system with Oxford Magnet.

\begin{figure*}
    \centering
      \includegraphics[width=18cm, height=14cm]{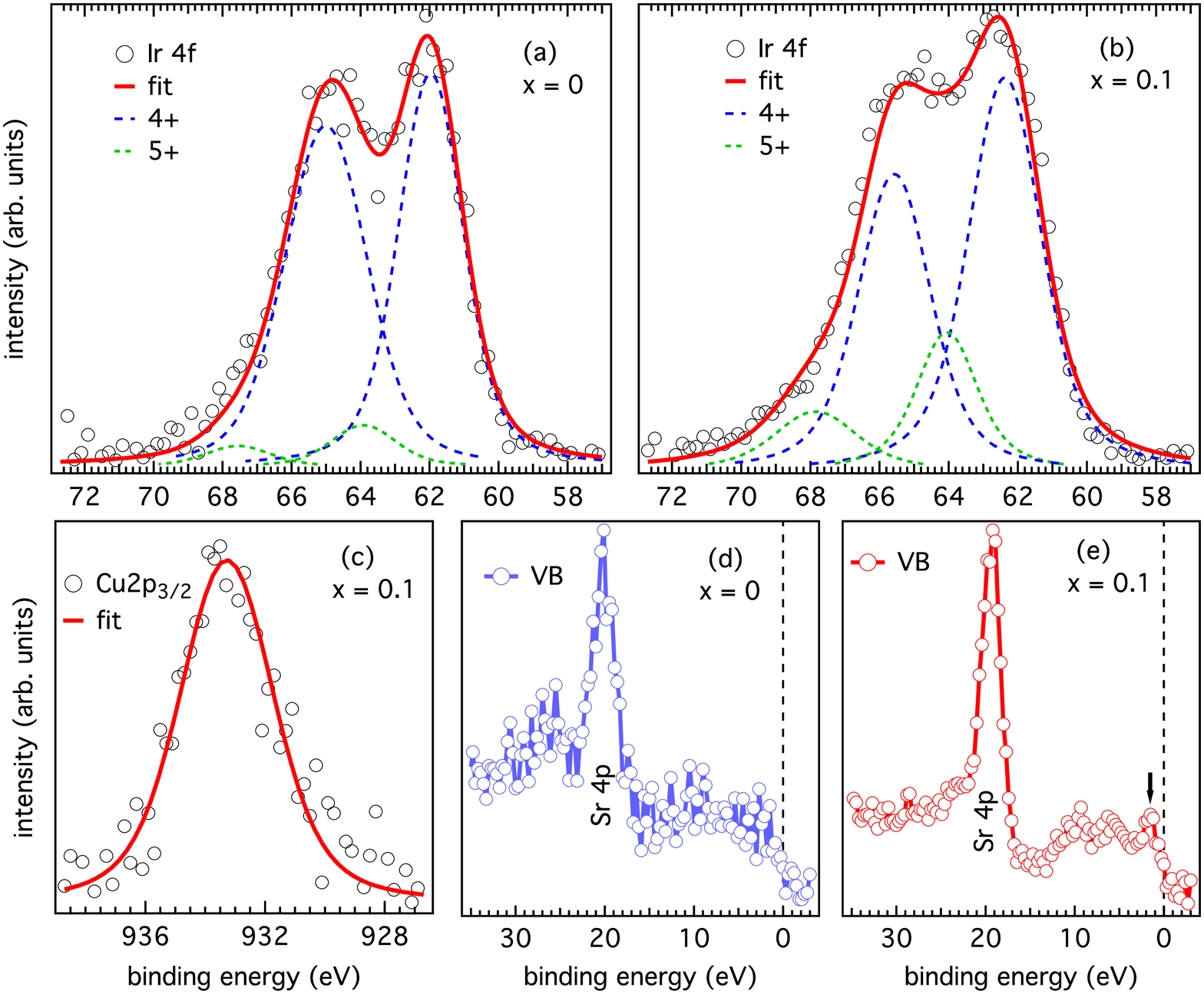}
    \caption{(color online) (a, b) The Ir 4$f$ core level spectra of $x =$ 0.0 and 0.1, respectively. Open black circles represent the experimental data and the red solid lines in (a) and (b) are the fitted envelope taking contributions of Ir$^{4+}$ and Ir$^{5+}$ components which are individually plotted in dashed blue and green colors, respectively. (c) The Cu 2$p_{3/2}$ core-level spectrum of $x =$ 0.1 sample. (d, e) The valence band spectra with Sr 4$p$ states for $x =$ 0.0 and 0.1 samples, respectively.}
    \label{fig:Fig2}
\end{figure*}

\section{Result and Discussion}
\subsection{Structural characterization}
Fig. 1a shows the XRD pattern of all the samples in present series. It is evident in figure that there is no modification in the XRD pattern in terms of peak position or impurity peak with Cu doping which primarily implies Cu substitution does not cause any major structural distortion in present series. Inset of Fig. 1a shows representative Rietveld refinement of XRD data for $x$ = 0.0 parent material which suggests reasonably good fitting. The Rietveld analysis shows the sample is in single phase and crystallizes in tetragonal structure with \textit{I4$_1$/acd} space group. Similarly, Rietveld refinement is performed for other samples in series with reasonably good fitting. While we find no structural phase transformation with Cu substitution, however, lattice parameters show a slight modification with $x$. Considering a slight mismatch in ionic radii of Ir$^{4+}$ (0.625 \AA) and Cu$^{2+}$ (0.73 \AA), changes in lattice parameters are quite expected. Fig. 1b shows evolution of lattice parameters $a$ and $c$ with Cu concentration. The figure shows $a$ increases and $c$ decreases with $x$. While this changes in lattice parameters is not that significant (below 0.4\%), the decreasing $c/a$ ratio implies tetragonal distortion is reduced with progressive substitution of Cu. Fig. 1c presents variation of basal plane Ir-O bond angel (left axis) showing $<$Ir-O1-Ir$>$ angel increases rather straightens with Cu concentration. The distortion or rotation of IrO$_6$ octahedra around $c$-axis ($\theta_{oct}$) plays a vital role in Sr$_2$IrO$_4$ which is believed to induce Dzyaloshinsky-Moriya (DM) type antisymmetric exchange interaction and weak ferromagnetic behavior. The right axis of Fig. 1c shows variation of $\theta_{oct}$ with $x$. It is evident in figure that $\theta_{oct}$ decreases from 11.3$^o$ at $x$ = 0.0 to 8.04$^o$ at $x$ = 0.2. This decrease of $\theta_{oct}$ is quite expected as in Cu-based layered material La$_2$CuO$_4$, the CuO$_6$ octahedra is rotated only by $\sim$ 3$^o$ which probably induces stronger AFM behavior and low magnetic moment in this material. Nonetheless, increase of both lattice parameter $a$ and bond angel $<$Ir-O1-Ir$>$ and decrease of $\theta_{oct}$ (Fig. 1) are consistent with each other.           

\subsection{X-ray photoemission spectroscopy study }
It is very important to know the charge state of constituent elements as the ionic state of Cu and the related modification of Ir charge state will influence the physical properties accordingly. In order to understand the cataionic charge distribution in Sr$_2$IrO$_4$ as well as its evolution with Cu substitution, we have performed the XPS measurements. Figs. 2a and 2b show the Ir-4\textit{f} core level spectra for representative $x$ = 0.0 and 0.1, respectively. We have performed detail analysis of XPS data using Voigt function, which includes a convolution of Gaussian and Lorentzian broadenings from different sources. The Ir-4\textit{f} core level spectra have been fitted with two peaks corresponding to Ir-4\textit{f}$_{7/2}$ and Ir-4\textit{f}$_{5/2}$ levels, observed at the binding energy of 62.0 and 65.0 eV, respectively. The origin of these two peaks is due to spin-orbital splitting with an energy difference of $\sim$ 3 eV. The detailed analysis of Ir-4\textit{f} confirms that the major contribution is coming from Ir$^{4+}$ states, but there is also small contribution from Ir$^{5+}$ ions in $x$ = 0.0 sample, see Fig. 2a. It is evident in Figs. 2a and 2b that for Ir$^{4+}$, the spin-orbit split peaks Ir-4\textit{f}$_{7/2}$ and Ir-4\textit{f}$_{5/2}$ arise at binding energy around 62.0 and 65.0 eV, respectively (as shown by the dashed blue lines).\cite{zhu} Similarly, Ir-4\textit{f}$_{7/2}$ and Ir-4\textit{f}$_{5/2}$ peaks for Ir$^{5+}$ are observed at binding energy 64 and 67.8 eV, respectively (as shown by the dashed green lines). The presence of Ir$^{5+}$ in parent $x$ = 0.0 sample (Fig. 2a) is quite interesting. Our analysis reveal that for Sr$_2$IrO$_4$ system, $\sim$ 96\% of the Ir cations are in Ir$^{4+}$ (5$d^5$) oxidation state whereas only $\sim$ 4\% of Ir are found in Ir$^{5+}$ (5$d^4$) state. The amount of Ir$^{5+}$ is not though substantial, and possibly arises due to non-stoichiometry of material. For doped sample in Fig. 2b, we observe a significant increase in the Ir$^{5+}$ intensity with respect to the Ir$^{4+}$ states, which qualitatively imply that the Ir$^{4+}$/Ir$^{5+}$ ratio decreases with Cu substitution. Indeed, our analysis show amount of Ir$^{4+}$ and Ir$^{5+}$ in $x$ = 0.1 sample is about 74 and 26\%, respectively. Note that with only 10\% of Cu substitution the amount of Ir$^{5+}$ has increased roughly by 22\%, if we consider similar amount of oxygen non-stiochiometry in the doped material as in parent material. This result clearly suggests Cu adopts charge state of Cu$^{2+}$ and each Cu converts two Ir$^{4+}$ ions into Ir$^{5+}$ ions.

To further comprehend the charge state of Cu in the present series, we have recorded the XPS spectra of Cu 2\textit{p}$_{3/2}$ core level, as presented in Fig. 2c for representative $x$ = 0.1 sample. The continuous red line represents the fitting of the XPS data, which found to be reasonably good. It shows that the Cu-2\textit{p}$_{3/2}$ spectra is centered around binding energy 933.4 eV confirming that the Cu is in Cu$^{2+}$ oxidation state.\cite{james} This result implies substitution of each Cu$^{2+}$ converts two Ir$^{4+}$ ions into Ir$^{5+}$ state. This finding is interesting as following spin-orbit coupling scheme, Ir$^{5+}$ is believed to be nonmagnetic which will definitely have large significance on the magnetic and transport properties in these materials. It is important to note that mostly the physical properties are controlled by the electronic states present near the Fermi level. Therefore, we recored XPS valence band (VB) spectra, as shown in Figs. 2d and 2e for $x$ = 0.0 and 0.1 samples, respectively, and discuss the changes in the states near the Fermi level. For both the samples, we observed the Sr-4$p$ peaks at around 20 eV binding energy. We now focus on the near Fermi level region in which there are no well defined peaks observed for $x$ = 0.0 sample (see Fig. 2d), which is consistent with the higher resistivity and its semiconducting nature. More interesting, with Cu substitution, we observe a relatively sharp peak close to the Fermi level, which is probably due to the contribution mostly from e$_g$ states of the Cu-3$d$, as marked by a black arrow in Fig. 2e. This indicates the larger density of states present at the Fermi level for $x$ = 0.1 sample and supports the observed significant decrease in the resistivity at room temperature with Cu substitution. It can be noted here that there is some fluctuation in XPS-VB data in Fig. 2d ($x$ = 0) which arises because of less amount of data collection time is given. Nonetheless, both figures show a comparative picture of electronic properties as Fig. 2e shows a prominent peak close to Fermi level in doped sample which is missing in parent material.

\begin{figure}
	\centering
		\includegraphics[width=8cm]{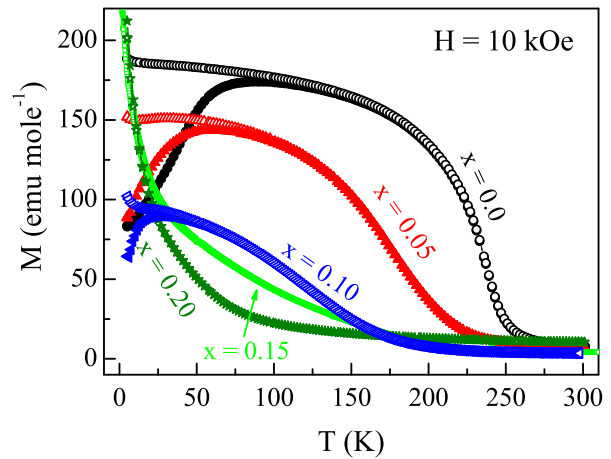}
	\caption{(color online) Magnetization measured in applied field of 10 kOe following ZFC and FC protocol are shown as function of temperature for Sr$_2$Ir$_{1-x}$Cu$_{x}$O$_4$ series.}
	\label{fig:Fig3}
\end{figure}

\begin{figure}
	\centering
		\includegraphics[width=8cm]{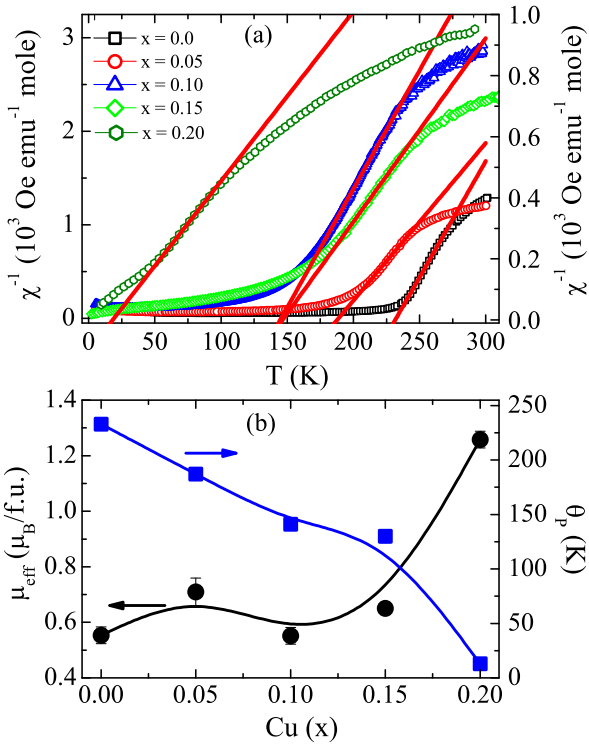}
	\caption{(color online) (a) shows temperature dependent inverse susceptibility ($\chi^{-1}$ = $(M/H)^{-1}$) as deduced from magnetization data shown in Fig. 3 for complete series of samples Sr$_2$Ir$_{1-x}$Cu$_{x}$O$_4$. The solid lines are due to fitting with Curie-Weiss law (Eq. 1). (b) shows effective magnetic moment $\mu_{eff}$ (left axis) and Curie temperature $\theta_P$ (right axis) as a function of Cu, respectively.}
	\label{fig:Fig4}
\end{figure}

\begin{figure}
	\centering
		\includegraphics[width=8cm]{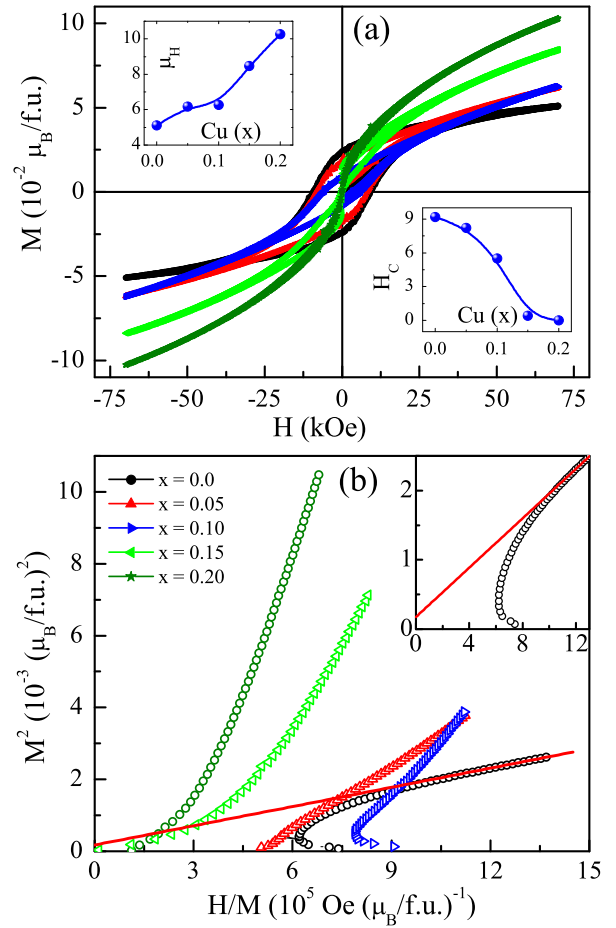}
	\caption{(color online) (a) shows isothermal magnetization as function of applied field up to $\pm$70 kOe collected at 5 K for series Sr$_2$Ir$_{1-x}$Cu$_{x}$O$_4$. Upper and lower inset show composition dependent moment $\mu_H$ at field 70 kOe and coercive field $H_c$, respectively. (b) shows Arrott plot ($M^2$ vs $H/M$) of $M(H)$ data shown in (a). Inset shows magnified view of Arrott plot for $x$ = 0.0 sample close to origin.}
	\label{fig:Fig5}
\end{figure}

\subsection{Magnetization study}
Fig. 3 shows the temperature dependent magnetization measured under zero-field cooled (ZFC) and field cooled (FC) protocol in an applied field of 10 kOe for Sr$_2$Ir$_{1-x}$Cu$_x$O$_4$ series. The nature of magnetism in Sr$_2$IrO$_4$ is rather interesting. In Sr$_2$IrO$_4$, Ir is assumed to adopt Ir$^{4+}$ charge state which gives 5$d^5$ electronic state. Due to high crystal field effect, all the five electrons will populate low energy $t_{2g}$ state realizing low-spin state. In strong SOC limit, it is believed that $t_{2g}$ electronic state is split into $J_{eff}$ = 3/2 quartet and 1/2 double. In this situation, 5$d^5$ state of Ir$^{4+}$ would give fully-filled $J_{eff}$ = 3/2 and half-filled $J_{eff}$ = 1/2 state. The spin-1/2 of Ir$^{4+}$ otherwise engage in AFM interaction, however, the distortion of IrO$_6$ octahedra around $c$-axis induces DM-type antisymmetric exchange interaction which consequently gives weak ferromagnetic behavior with $T_c$ $\sim$ 225 K.\cite{fujiyama,imtiaz1} Fig. 3 depicts $M(T)$ for $x$ = 0.0 material shows sudden rise below around 240 K which marks PM to (weak) FM transition. On further cooling below $\sim$ 95 K, $M_{ZFC}(T)$ data show a decrease and open a gap between $M_{FC}$ and $M_{ZFC}$ which has recently been shown to arise due to prominent magneto-structural coupling where the spin and lattice degrees of freedom are coupled and suffer staggered rotation with structural evolution with temperature.\cite{imtiaz} For the doped materials, along with the decreasing moment the PM-FM phase transition temperature $T_c$ decreases also with $x$. This weakening of FM state is clearly evident for samples with $x$ = 0.15 and above where $T_c$ seems to disappear and the $M(T)$ plot looks more like PM samples. This is quite interesting as Cu$^{2+}$ (3$d^9$) which is magnetically active with spin-1/2, substituted for Ir$^{4+}$ ($J_{eff}$ = 1/2) will not act for site dilution, although Cu$^{2+}$ will create double amount of Ir$^{5+}$ ($J_{eff}$ = 0) which is nonmagnetic and will act for site dilution. However, active $d$-orbitals for former and later cations are $e_g$ and $t_{2g}$, respectively which are orthogonal so its magnetic compatibility to be considered. The suppression of long-range magnetic ordering in Sr$_2$IrO$_4$ has also been observed doping with other elements which is discussed later.

To look into the magnetic state in further detail, we have plotted inverse susceptibility ($\chi^{-1}$) as a function of temperature for this series in Fig. 4a. For $x$ = 0.0 parent material, $\chi^{-1}(T)$ shows linear behavior above $T_c$, however, in high temperature above 280 K, the $\chi^{-1}$(T) deviates from linearity. For doped samples, we also observe similar behavior in PM state though $\chi^{-1}$(T) becomes more linear with $x$. The Sr$_2$IrO$_4$ is known to be structurally layered material with $n$ = 1 in Ruddlesden-Popper series Sr$_{n+1}$Ir$_n$O$_{3n+1}$ where SrIrO$_3$ layer is separated by SrO showing 2-dimensional behavior. In fact, 2-dimensional like anisotropic magnetic interaction for in-plane and out-plane has been shown using resonant magnetic x-ray diffuse scattering experiment with exchange coupling constant $J$ $\sim$ 0.1 and 10$^{-6}$ eV, respectively.\cite{fujiyama} This study further shows in-plane exchange interaction survives at least for $\sim$ 25 K above $T_N$ = 228.5 K. We find that deviation from linearity in our $\chi^{-1}(T)$ for Sr$_2$IrO$_4$ starts at $\sim$ 270 K (Fig. 4a) which is in qualitative agreement with other microscopic study. We believe that this change in slope in $\chi^{-1}(T)$ well within PM state is linked to anisotropic magnetic interaction. However, the persistence of spin correlation at temperatures much higher than the magnetic ordering temperature appears to be typical feature of layered materials as seen for La$_2$CuO$_4$,\cite{keimer} or in case of iron pnictides.\cite{klingeler} In this sense, increase of linearity of $\chi^{-1}(T)$ with $x$ implies spin interaction is weakened with Cu doping. 

The straight lines in Fig. 4a are due to fitting with Curie-Weiss (CW) law,

\begin{eqnarray}
 \chi = \frac{C}{T - \theta_P}
\end{eqnarray}

where $C$ (= N$_A$$\mu^2$$_{eff}$/3k$_B$) is the Curie constant, $\mu_{eff}$ is the effective PM moment and $\theta_P$ is the Curie temperature. Fig. 4a shows just above $T_c$, $\chi^{-1}(T)$ data can be reasonably fitted with Eq. 1. Using the fitted parameter $C$, we have calculated $\mu_{eff}$ for all the samples. Fig. 4b shows composition dependent $\mu_{eff}$ (left axis) and $\theta_P$ (right axis). We calculate $\mu_{eff}$ = 0.553 $\mu_B$/f.u for $x$ = 0.0 which appears much lower than the calculated value (for spin-only $g\sqrt{S(S+1)}\mu_B$) 1.72 $\mu_B$/f.u for spin-1/2 material. The $\mu_{eff}$ shows a slight increase in value for $x$ = 0.05 which can be due to active magnetic interaction of Cu in low doping regime. With increasing $x$, $\mu_{eff}$ though initially shows some fluctuation, but for the highest doped sample with $x$ = 0.2, where $M(T)$ is more like PM-type, it shows steep increase yielding the value 1.26 $\mu_B$/f.u. While both Ir$^{4+}$ and Cu$^{2+}$ are spin-1/2 element, Ir$^{5+}$ is believed to be non-magnetic ($S$ = 0). Therefore, for $x$ amount of Cu$^{2+}$ which generates 2$x$ of Ir$^{5+}$, the average effective moment is $\mu_{eff}$ = $\sqrt{(1 - 3x)(^{Ir^{4+}}\mu_{eff})^2 + 2x(^{Ir^{5+}}\mu_{eff})^2 + x(^{Cu^{2+}}\mu_{eff})^2}$, which implies that substitution of Cu would decrease $\mu_{eff}$, provided spins are noninteracting in PM state. On contrast, sudden increase of $\mu_{eff}$ at higher value of $x$ suggests Cu destabilizes the magnetic interaction in Sr$_2$IrO$_4$ and promotes the PM state. The composition dependent $\theta_P$ shown in Fig. 4b shows $\theta_P$ for parent Sr$_2$IrO$_4$ is as high as 233 K closely matching its $T_c$ = 225 K, but the $\theta_P$ continuously and sharply decreases with $x$. Even though, $\theta_P$ remains positive for highest $x$ = 0.2, its value of 13 K shows strength of magnetic interaction is substantially weakened (Fig. 4b). We have not observed any peak or any bifurcation in M$_{ZFC}$ and M$_{FC}$ data in low temperature for $x$ = 0.2, but this small and finite value of $\theta_P$ prpbably causes some nonlinearity in $\chi^{-1}(T)$. Nonetheless, Cu$^{2+}$substituted for Ir$^{4+}$ has substantial influence on magnetic behavior of Sr$_2$IrO$_4$.

\begin{figure}
	\centering
		\includegraphics[width=8cm]{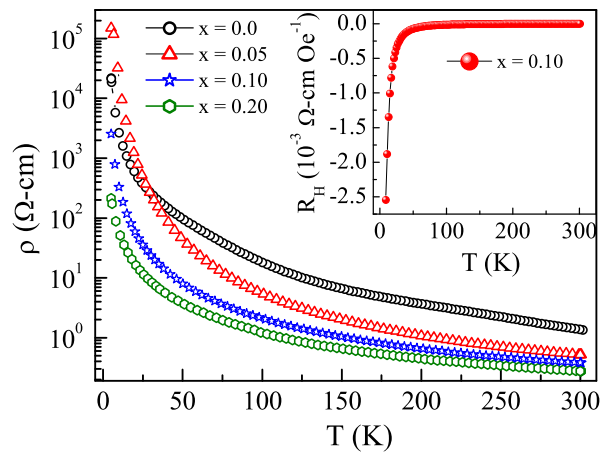}
	\caption{(color online) Temperature dependent resistivity are shown for Sr$_2$Ir$_{1-x}$Cu$_{x}$O$_4$ series. Inset shows the hall coefficient $R_H$ as a function of temperature in an applied field of 10 kOe.}
	\label{fig:Fig6}
\end{figure}

Magnetic-field dependent magnetization $M(H)$ measured at 5 K in the range of $\pm$ 70 kOe are shown in Fig. 5a. The $M(H)$ for Sr$_2$IrO$_4$ shows large opening at low temperatures with coercive field $H_c$ $\sim$ 9370 Oe. With Cu substitution, major observations are moment $\mu_H$ at highest field increases and $H_c$ substantially decreases where for $x$ = 0.2 material, we find $H_c$ $\sim$ 100 Oe. The variation of $\mu_H$ and $H_c$ with Cu composition are shown in upper and lower inset of Fig. 5a, respectively. In Fig. 5b, we have shown Arrott plot of $M(H)$ data. Arrott plot ($M^2$ vs $H/M$) offers an effective tool to understand the nature of  magnetic state as the positive intercept on $M^2$ axis due to straight line fitting in high field regimes of Arrott plot implies spontaneous magnetization or FM state. For Sr$_2$IrO$_4$, Fig. 5b shows very small positive intercept (magnified view is shown in inset) which indicates weak ferromagnetism in this material, as expected. Cu substitution gives negative intercept in Arrott plot, where for higher doping intercept is highly negative which excludes the possibility of any FM behavior or rather magnetic ordering in present case. The increase of $\mu_{eff}$ and simultaneous decrease of $H_c$ are indicative of weakening of (AFM type) magnetic ordering which sets the spins free.

\begin{figure}
	\centering
		\includegraphics[width=8cm]{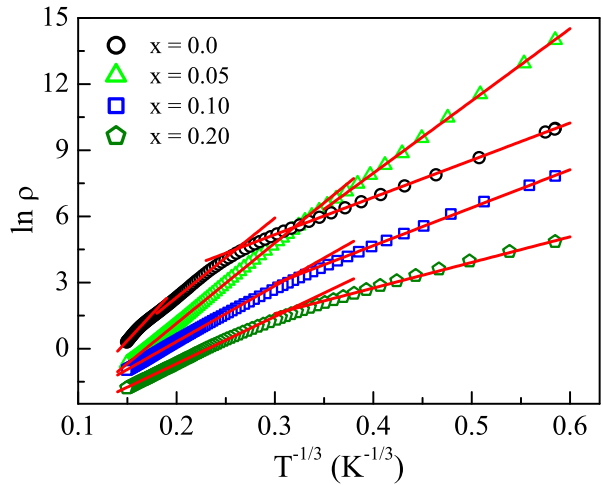}
	\caption{(color online) Logarithm of resistivity with $T^{-1/3}$ is shown for Sr$_2$Ir$_{1-x}$Cu$_{x}$O$_4$. The solid lines are the least-square fitting due to two dimensional Mott's variable-range-hoping model following Eq. 2 in different temperature ranges. The data for $x$ = 0.20 data has been shifted downward for clarity).}
	\label{fig:Fig7}
\end{figure}

\subsection{Electrical transport}
The effect of Cu substitution on charge transport is shown in Fig. 6 which depicts resistivity $\rho$ as a function of temperature for Sr$_2$Ir$_{1-x}$Cu$_{x}$O$_4$ series. The Sr$_2$IrO$_4$ is a well know insulator where the intriguing insulating state is believed to occur due to opening a gap in $J_{eff}$ = 1/2 electronic state.\cite{kim1} Recently, it has been shown some correlation between structural, magnetic and electronic transport behavior in Sr$_2$IrO$_4$. The Cu$^{2+}$ is similarly spin-1/2 system but its level of $d$-orbial filling is different compared to Ir$^{4+}$ and acts as electron doping with total 3$d^9$ electrons. On the other hand, it will not only tune the SOC and $U$ parameters with its 3$d$ character but it will also create double amount of Ir$^{5+}$ which has empty $J_{eff}$ = 1/2 doublet state. 

The parent Sr$_2$IrO$_4$ shows highly insulating behavior where the resistivity increases by roughly four orders going to low temperatures in agreement with earlier studies. In doped materials, resistivity decreases continuously with Cu in higher temperature regime but in low temperature we observe that $\rho(T)$ initially increases for $x$ = 0.05 then decreases. The overall system remains insulating till highest doping level ($x$ = 0.2). The decrease of resistivity is, however, very prominent in low temperatures. This decrease of resistivity could be caused by many factors such as, effect of electron doping through Cu$^{2+}$ which would modify the Fermi level, tuning of $J_{eff}$ states through modification of SOC and $U$, conversion of Ir$^{5+}$ states which will have empty $J_{eff}$ = 1/2 that would promote hopping of electrons. Hall measurements on Sr$_2$IrO$_4$, as reported by Klein \textit{et al},\cite{klein} have shown charge carriers in this material are hole type i.e., $p$-type characteristics of this material. We have done temperature dependence of Hall measurements in applied field of 10 kOe for a selective doped sample i.e., with $x$ = 0.1. Temperature dependence of Hall coefficient $R_H$ (=1/$ne$) is shown in inset of Fig. 6. Negative value of $R_H$ indicates charge carriers are basically electron type where this conversion from hole- to electron-type has occurred through doping of electron through Cu$^{2+}$ substitution. 

\begin{table}
\caption{\label{label} Temperature range and fitting parameter $T_0$ obtained from fitting of Eq. 2 in Fig. 7 are given for Sr$_2$Ir$_{1-x}$Cu$_x$O$_4$ series}
\begin{ruledtabular}
\begin{tabular}{ccc}
Sample &Temperature range (K) &$T_0$ (K)  \\
Sr$_2$Ir$_{1-x}$Cu$_x$O$_4$ &  &(10$^5$)\\
\hline
x = 0.0		&300 - 240  &1.44(1) \\
					&240 - 70 	&0.46(8) \\
					&40 - 5 		&0.04(8) \\
\hline
x = 0.05	&300 - 35 &0.48(6) \\
					&30 - 5  	&0.35(2) \\
\hline
x = 0.10	&300 - 33 &0.16(1)  \\
					&25 - 5 &0.05(1) \\
\hline
x = 0.20  &300 - 30 &0.11(4) \\
					&20- 5 	&0.02(1) \\
\end{tabular}
\end{ruledtabular}
\end{table}

In an attempt to understand the nature of charge conduction in present series, we found that resistivity can be best explained using thermally activated conduction mechanism described by 2-dimensional(D) Mott-variable range hoping (VRH) model,\cite{mott}

\begin{eqnarray}
\rho = \rho_0 \exp\left[\left(\frac{T_0}{T}\right)^{1/3}\right]	
\end{eqnarray}

where $T_0$ is the characteristic temperature and can be expressed as:

\begin{eqnarray}
	T_0 = \frac{21.2}{k_BN(E_F)\xi^3}
\end{eqnarray} 

where k$_B$ is the Boltzmann constant, N(E$_F$) is the density of states (DOS) at Fermi level and $\xi$ is the localization length. 

Here to be noted that we have previously shown that charge conduction in Sr$_2$IrO$_4$ follows Mott's 2D variable range hopping model in three different temperature regimes which are related to magnetic and or structural state in this material.\cite{imtiaz} Moreover, 2D nature of charge transport is in agreement with layered structure of Sr$_2$IrO$_4$. Taking this as a guideline, we have analyzed the $\rho(T)$ data for all the compositions using Eq. 2. Fig. 7 shows straight line fitting of $\rho(T)$ data following Eq. 2 where for parent Sr$_2$IrO$_4$, the data can be fitted in three distinct temperature regimes. The temperature regimes and obtained fitting parameters $T_0$ are given in Table I. The temperature regimes mark the change of magnetic state (Fig. 3) and structural parameters.\cite{imtiaz} For the doped samples, however, we can fit the data only in two distinct temperature regimes (Fig. 7).The temperatures ranges and the parameter $T_0$ are given in Table I. This could be understood as magnetic transitions in doped materials are broadened which has led to single temperature fitting behavior in high temperatures. Nonetheless, change of slope in Fig. 7 for $x$ = 0.2 compound is intriguing as magnetization data show no clear magnetic transition. This could be due to existence of short-range type of magnetic ordering as evident in $\chi^{-1}(T)$ where a slope change is observed at low temperatures (Fig. 4a).
 
\begin{figure}
	\centering
		\includegraphics[width=8cm]{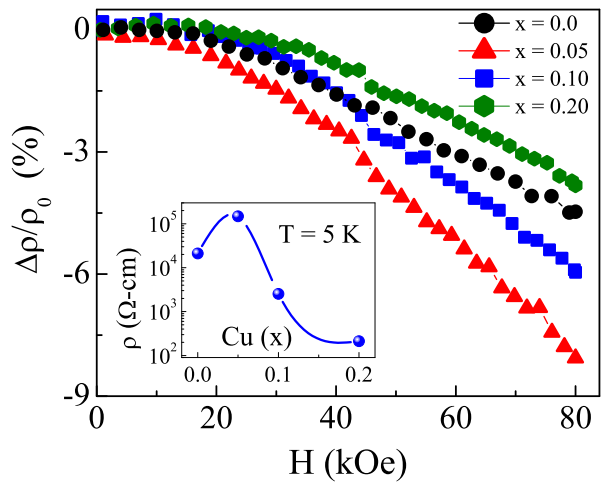}
	\caption{(color online) Magnetoresistance measured at 5 K in magnetic field up to 80 kOe are shown for Sr$_2$Ir$_{1-x}$Cu$_x$O$_4$ series. Inset shows the variation of resistivity with doping concentration $x$ at 5 K.}
	\label{fig:Fig8}
\end{figure}
 
Table I further indicates that parameter $T_0$ shows lower value both with lowering the temperature as well as with increasing the Cu concentration. For Sr$_2$IrO$_4$, we have previously shown that change of $T_0$ with lowering of temperature can be explained through increase of localization length $\xi$ as N(E$_F$) would unlikely modify for an insulating sample (Eq. 3).\cite{imtiaz} However, the modification of $T_0$ with $x$ it can not be straightforwardly explained as increase of both N(E$_F$) and $\xi$ parameter can contribute to decrease of $T_0$. This is a quite likely phenomenon as modification of electronic conductivity (Fig. 6), structural parameters (Fig. 1) and magnetic state (Fig. 3) with Cu indicates both N(E$_F$) and $\xi$ would possibly change with $x$. Nonetheless, large change in conductivity in Sr$_2$Ir$_{1-x}$Cu$_x$O$_4$ series imply that N(E$_F$) has major contributions for change of $T_0$.

For further understanding of electronic transport behavior, we have measured electrical resistivity in presence of magnetic field at 5 K. Fig. 8 shows the change of resistivity which is commonly known as magnetoresistance (MR) and calculated as, $\Delta \rho/\rho(0)$ = $\left[\rho(H) - \rho(0)\right]/\rho(0)$ for Sr$_2$Ir$_{1-x}$Cu$_x$O$_4$ series. All the samples show negative MR i.e., conductivity increases in presence of magnetic field. In picture of variable range hopping, negative MR is considered to arise due to `weak localization' phenomenon, induced by quantum interference effect.\cite{nguyen,sivan} Recently, we have shown that parent Sr$_2$IrO$_4$ exhibits negative MR and its quadratic field dependence which is considered as typical feature of weak localization effect.\cite{imtiaz} Fig. 8 shows evolution of MR with field is not linear where the resistivity start to decrease above around 20 kOe. MR value is not, however, substantial and Sr$_2$IrO$_4$ shows $\sim$ 4.5\% MR in field as high as 80 kOe. As Cu is introduced, general feature of MR with field remains similar but its magnitude initially increases showing highest value for $x$ = 0.05, and then decreases. While exact reason for this anomalous change of MR is not very clear, but in lower concentration of Cu i.e., for $x$ = 0.05 we have already seen some anomaly in magnetic moment, $\mu_{eff}$ and $\mu_H$ in Figs. 4b and 5a, respectively. Moreover, it is evident in Fig. 6 that resistivity at low temperatures initially increases with $x$. In inset of Fig. 8, we have shown composition dependent resistivity value at 5 K in zero field, showing a nonmonotonic behavior. While the structural parameters do not exhibit any anomalous behavior for low doped samples (Fig. 1), this anomaly in magnetic moment as well as in electronic transport behavior for $x$ = 0.05 samples appears to be correlated effect. From these observations we speculate that Cu in lower concentration induces magnetic moment which perhaps inhibits the hopping of charge carriers (VRH), hence resistivity increases at low temperatures. The application of magnetic field tends to orient the moments which may be helpful for hopping mechanism, therefore negative MR increases. In higher concentration of Cu, along with increase of electrons the Ir$^{5+}$ also increases which can reverse the situation.

\begin{figure}
	\centering
		\includegraphics[width=8cm]{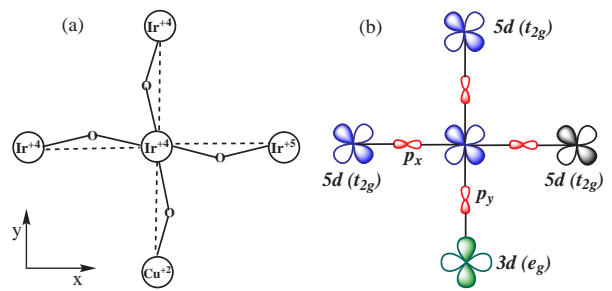}
	\caption{(color online) (a) Visualization of possible arrangement of Ir$^{4+}$, Ir$^{5+}$ and Cu$^{2+}$ ions. (b) The schematic arrangement of Cu-3\textit{d} and Ir-5\textit{d} orbitals are shown.}
	\label{fig:Fig9}
\end{figure}

\subsection{Possible ionic arrangement and interaction model}
Our studies show that Cu substitution in Sr$_2$Ir$_{1-x}$Cu$_x$O$_4$ series has prominent effects on magnetism and electronic transport behavior where the magnetic state is largely weakened giving PM-like behavior for $x$ = 0.2 material and the charge conductivity overall increases. The similar weakening or suppression of magnetic ordering and insulating state has previously been observed in Sr$_2$IrO$_4$ with different chemical substitution. For Sr-site doping with La$^{3+}$, which acts for electron doping creating equivalent amount of Ir$^{3+}$ (5$d^6$), the magnetic state is completely suppressed with $\sim$3\% of doping concentration in (Sr$_{1-x}$La$_x$)$_2$IrO$_4$.\cite{chen,gretarsson} In case of Ir-site doping, results with different elements are different. For instance, Tb$^{4+}$ substitution which is believed to change the relative strength of SOC, crystal field effect and Hund's coupling with its 4$f^7$ character causes complete suppression of magnetic state with mere 3\% of doping concentration.\cite{wang} While the charge state of Rh is debated (Rh$^{3+}$ or Rh$^{4+}$ which would induce hole- or isoelectronic-doping, respectively), its substitution in Sr$_2$Ir$_{1-x}$Rh$_x$O$_4$ has shown suppression of ordered magnetic and insulating state above the critical doping concentration of $\sim$ 17\%.\cite{clancy,qi,cao1} Similarly, Ru substitution has caused suppression of magnetic ordering in Sr$_2$IrO$_4$ with nearly 30 to 50\% of Ru$^{4+}$ doping level.\cite{calder,yuan} 

In present Sr$_2$Ir$_{1-x}$Cu$_x$O$_4$ series, both Ir$^{4+}$ and Cu$^{2+}$ is spin-1/2 element and magnetically active, however, respective orbitals are not symmetric. For instance, in strong SOC limit Ir$^{4+}$ has half-filled $J_{eff}$ = 1/2 state ($t_{2g}$) while the single spin in Cu$^{2+}$ populate $d_{x^2-y^2}$ orbital ($e_g$). The incompatibility of magnetic interaction between transition metals with orthogonal $d$ character has been discussed for spin chain material Sr$_3$MIrO$_6$ (M = Ni, Co).\cite{ou} In addition, Cu$^{2+}$ substitution will create double amount of Ir$^{5+}$ (5$d^4$) which are expected to be nonmagnetic considering four electrons will fully fill $J_{eff}$ = 3/2 quartet state. In transition metal oxides, metal octahedra are generally corner shared and the magnetic interactions are mediated through O-2\textit{p} orbitals. While ionic distribution is quite random, based on present scenario the magnetic interaction scheme is depicted in Fig. 9. The Ir$^{5+}$ being nonmagnetic, exchange interaction through Ir$^{4+}$-O$^{2-}$-Ir$^{5+}$ channel is very unlikely, however, the same through Ir$^{4+}$-O$^{2-}$-Ir$^{4+}$ and Ir$^{4+}$-O$^{2-}$-Cu$^{2+}$ channels are of interest. Note, that in trivalent doped (Sr$_{1-x}$La$_x$)$_2$IrO$_4$, where La$^{3+}$ doping creates Ir$^{3+}$ (5$d^6$) ions which are again nonmagnetic (fully filled $J_{eff}$ = 3/2 quartet and $J_{eff}$ = 1/2 doublet states), the long-range magnetic ordering is suppressed with only $\sim$3\% of La doping.\cite{chen} In contrast, magnetism survives till higher doping level ($\sim$20\%) of Cu$^{2+}$ substitution at Ir-site in present study. Similar results are also evident for Rh$^{3+}$ and Ru$^{4+}$ where magnetism is totally suppressed with $\sim$ 17 and 30\% of doping concentration, respectively.\cite{clancy,calder} Given that Rh$^{3+}$ and Cu$^{2+}$ creates nonmagnetic Ir$^{5+}$, both has similarity as they create double amount of nonmagnetic sites; Rh$^{3+}$ and Ir$^{5+}$ in former case and Ir$^{5+}$ and Ir$^{5+}$ in later case. Of course, Cu$^{2+}$ being magnetic, effective magnetic exchange would be different and complex compared to Rh$^{3+}$. For 20\% of Cu, magnetic interaction is substantially weakened and is almost suppressed though we find some trace of magnetism. Thus, like Rh$^{3+}$ doping,\cite{clancy} percolation picture to explain the suppression of magnetism apparently looks valid for Cu$^{2+}$ also. The evolution of magnetic and transport properties with $x$ in Sr$_2$Ir$_{1-x}$Cu$_x$O$_4$ is quite intriguing and needs to be investigated using microscopic probe as well as theoretical models. 

\subsection{Conclusion}
In conclusion, we have studied the structural, magnetic and electronic transport properties in series of polycrystalline samples Sr$_2$Ir$_{1-x}$Cu$_x$O$_4$ with $x$ $\leq$ 0.2. Structural investigation shows while though system retains its original structural symmetry, the unit cell parameters modify with Cu. In particular, both $c/a$ ratio and distortion of IrO$_6$/CuO$_6$ octahedra decreases with $x$. The XPS results reveal that doped Cu$^{2+}$ converts double amount of Ir$^{4+}$ in to Ir$^{5+}$ and sharpens the valence band spectra at Fermi level. Substitution of Cu$^{2+}$ introduces electron doping which is evident in Hall measurements. Long-range magnetic ordering in parent material Sr$_2$IrO$_4$ is destabilized with Cu and in highest doped sample i.e., in Sr$_2$Ir$_{0.8}$Cu$_{0.2}$O$_4$ the magnetic data show paramagnetic-like behavior. Although resistivity decreases but the system does not become metallic with available 20\% of Cu doping. The charge conduction follows Mott's 2D VRH model, but interestingly validity of this model changes with temperature and appears to follow the magnetic behavior. A weak negative MR has been observed at low temperature which is considered to be a signature of weak localization effect in spin-orbit coupled system. The evolution of MR shows nonmonotonic behavior with Cu substitution, rather its appears to follow the variation of resistivity with $x$. 

\section{Acknowledgment} 
We acknowledge UGC-DAE CSR, Indore and Alok Banerjee, Rajeev Rawat and Archana Lakhani for magnetization, resistivity and Hall measurements. We also thank Kranti Kumar and Sachin Kumar for the helps in measurements. INB acknowledges CSIR, India for fellowship.

\end{document}